# Resonant soft X-ray Raman scattering of NiO

M. Magnuson, S. M. Butorin, A. Agui and J. Nordgren

*Department of Physics, Uppsala University, P. O. Box 530, S-751 21 Uppsala, Sweden*

**Abstract**

Resonant soft X-ray Raman scattering measurements on NiO have been made at photon energies across the Ni 2*p* absorption edges. The details of the spectral features are identified as Raman scattering due to *d−d* and charge-transfer excitations. The spectra are interpreted within the single impurity Anderson model, including multiplets, crystal-field and charge-transfer effects. At threshold excitation, the spectral features consists of triplet-triplet and triplet-singlet transitions of the $3d^8$ configuration. For excitation energies corresponding to the charge-transfer region in the Ni 2*p* X-ray absorption spectrum of NiO, the emission spectra are instead dominated by charge-transfer transitions to the $3d^9\underline{L}^{-1}$ final state. Comparisons of the final states with other spectroscopical techniques are also made.

## 1 Introduction

In recent years, the interest in transition metal compounds has grown significally in several branches of solid state physics [1]. Some of these systems show a variety of exotic properties such as high-$T_c$ superconductivity, giant magnetoresistance and insulating behavior. NiO has often been considered as a model system for studying strong electron correlations and the applicability of theoretical models. Ordinary band-structure calculations performed on NiO is known to show rather large discrepancies for the band-gaps and the existence of satellite structures in experimental spectra [2, 3]. However, it has been shown that self-energy treatments and other methods improve excitation spectra and their satellite structures as well as the band gap [4, 5]. In contrast to the free ion, the degeneracy of the 3*d* states is partially lifted so that the $t_{2g}$ and $e_g$ states of the $Ni^{2+}$ ion are energetically separated by the $O_h$ crystal-field splitting arising from the six octahedrally surrounding $O^{2-}$ ligands. The ground state of NiO possesses a spin-triplet $^3A_{2g}(t_{2g})^6(e_g)^2$ symmetry and the hybridization of the metal ion states with the O 2*p* ligand states result in substantial charge transfer character. In the localized description of the 3*d* electrons, which is known to describe excitation spectra rather well, the $3d^8$, $3d^9\underline{L}^{-1}$ and $3d^{10}\underline{L}^{-2}$ configurations ($\underline{L}^{-1}$ denotes a hole in the O 2*p* ligand band) are energetically split up by the ligand field [6]. This implies several triplet and spin-flip singlet excited states within $\sim$ 3 eV from the ground state. Due to ionic-lattice vibrations, direct *d−d* transitions which are normally dipole forbidden become weakly allowed and have been identified in optical absorption spectroscopy (OAS) [7, 8]. These faint excitations can also be studied with electron energy loss spectroscopy (EELS) [9, 10] since the dipole selection rules are relaxed at low electron energies but these measurements are very surface sensitive. While detailed resonant photoemission spectroscopy (RPES) measurements aimed at the 2*p* and 3*p* core levels [11, 12, 13] confirm the charge-transfer character of NiO, the spectra are mixed with the O 2p band and it is not clear to which extent the $3d^7$ and the screened $3d^8\underline{L}^{-1}$ final states reflects the electronic structure of the ground state [2]. Although transition metal oxides are known to exhibit charge-transfer insulator properties, this important mechanism is not yet fully understood.

In this work we investigate the electronic structure of NiO using resonant inelastic X-ray scattering (RIXS) spectroscopy with selective excitation energies around the Ni 2p thresholds. This technique is more bulk sensitive than RPES and each atomic element can be probed separately by tuning the excitation energy to the appropriate core edge. The d−d excitations to the various excited states which can be studied in terms of loss structures become fully allowed due to the core-hole assisted excitation-deexcitation dipole transitions [14]. The 2p spin-orbit coupling also allows Hund's rule electron exchange scattering of triplet-singlet spin-flip transitions. When the excitation energy is tuned to the different features in the absorption spectrum, the RIXS spectra of NiO are found to exhibit resonant energy-loss structures due to both d−d excitations of Raman scattering and charge-transfer excitations of $3d^9\underline{L}^{-1}$ final state character. The RIXS spectra are interpreted with support of Anderson impurity calculations using the same set of parameters as in X-ray absorption. Although the final states of RIXS are necessarily not the same as in other spectroscopical techniques, it is useful to compare the energy positions of the peaks and the validity of localized calculations for interpretation of the spectra.

## 2  Experimental Details

The measurements were performed at beamline BW3 at HASYLAB, Hamburg, using a modified SX700 monochromator [15]. An XAS spectrum at the Ni 2p edges was obtained in total electron yield (TEY) by measuring the sample drain current. The Ni $L_{2,3}$ RIXS spectra were recorded using a high-resolution grazing-incidence grating spectrometer with a two-dimensional position-sensitive detector[16]. During the XAS and RIXS measurements at the Ni 2p edges, the resolutions of the beamline monochromator were about 0.4 eV and 0.5 eV, respectively. The RIXS spectra were recorded with a spectrometer resolution better than 0.5 eV.

The measurements at the Ni 2p thresholds were performed at room temperature with a base pressure lower than $5\times10^{-9}$ Torr. During the absorption measurements, the NiO(100) single crystal was oriented so that the photons were incident at an angle of about 90o with respect to the sample surface. During the emission measurements, the angle of incidence was about 20o in order to minimize self-absorption effects[17]. The emitted photons were always recorded at an angle, perpendicular to the direction of the incident photons, with the polarization vector parallel to the horizontal scattering plane.

## 3  Calculational Details

The Ni 3d→2p RIXS spectra of NiO were calculated as a coherent second-order optical process including interference effects using the Kramers-Heisenberg formula [18]:

$$I(\Omega,\omega)=\sum_f \left| \sum_i \frac{\langle f|D_q|i\rangle \langle i|D_q|g\rangle}{E_g+\Omega-E_i-i\Gamma_i/2} \right|^2$$

$$\times \delta(E_g + \Omega - E_f - \omega).$$

The $\Omega$ and $\omega$ denotes the excitation and emission energies, the $|g>$, $|i>$ and $|f>$ are the ground, intermediate and final states with energies $E_g$, $E_i$ and $E_f$. $\hat{D}$ is the dipole operator and $\Gamma_i$ is the full width half maximum (FWHM) of the Lorenzian of each intermediate state representing the lifetime broadening which interfere between the different intermediate states. The values of the $\Gamma_i$s used in the calculations were 0.6 eV and 0.8 eV for the $L_3$ and $L_2$ thresholds, respectively[19]. The Slater integrals, describing $3d$–$3d$ and $3d$–$2p$ Coulomb and exchange interactions, and spin-orbit constants were obtained by the Hartree-Fock method[20]. The effect of the configurational dependent hybridization was taken into account by scaling the Slater integrals to $F^k(3d3d)$ 80%, $F^k(2p3d)$ 80% and $G^k(2p3d)$ 80%. The ground state of the Ni$^{2+}$ ion has $^3A_{2g}$ character in $O_h$ symmetry. In order to take into account the polarization dependence and exchange interactions, the calculations were made in the $C_{4h}$ basis set at 0 K. Two configurations were considered: $3d^8$ and $3d^9\underline{L}^{-1}$ for the initial and final states, and $2p^53d^9$, $2p^53d^{10}\underline{L}^{-1}$ for the intermediate states. The weights of the $3d^8$ and $3d^9\underline{L}^{-1}$ configurations in the ground state were 82% and 18%, respectively. The contribution of the $3d^{10}\underline{L}^{-2}$ configuration in the ground state was neglected here since its weight was estimated to be only in the order of ~ 0.5 % in prior studies[21].

The single impurity Anderson model (SIAM) [22] with full multiplet effects was applied to describe the system. The crystal field and exchange-interactions were taken into account by using a code by Butler [23] and the charge-transfer effect was implemented with a code by Thole and Ogasawara [24]. The SIAM parameters were chosen to reproduce the experiment as follows: the charge-transfer energy $\Delta$, defined as the energy difference between the center-of-gravity between the $3d^8$ and the $3d^9\underline{L}^{-1}$ configurations was 3.5 eV, the crystal-field splitting $10Dq$ was set to 0.5 eV and the inter-atomic exchange field was applied in the direction of the polarization vector of the incoming photons. The calculations were made for the same geometry as the experimental one, with the scattering angle between the incoming and outgoing photons fixed to 90$o$. The polarization dependence can be understood from a group theoretical consideration where according to the Wigner-Eckart theorem, the transition matrix elements are described by the Clebsch-Gordan coefficients in the $C_{4h}$ symmetry. The shape of the oxygen valence band, was appoximated by a function describing a circle with a width of 5.0 eV. The hybridization strength between the Ni $3d$ band and the O $2p$ band where $V_{e_g}$ represents the hopping for the Ni$^{2+}$ $e_g$ orbitals was taken to be 2.2 eV and 1.8 eV, for the ground and intermediate states, respectively. The smaller value of the hybridization strength of the intermediate states is due to the configurational dependence [25, 26]. The value of the hybridization strength for the Ni$^{2+}$ states of $t_{2g}$ symmetry $V_{t_{2g}}$ was taken as half of the value for the $e_g$ states $V_{e_g}$ which has previously been shown to be a reasonable empirical relation [27]. The interatomic exchange interactions which correspond to strong effective magnetic fields were taken into account using a mean-field theory [28]. The parameters used in the calculations are summarized in Table I.

# 4 Results and Discussion

Figure 1 shows a set of RIXS spectra of NiO recorded at different excitation energies at the Ni *2p$_{3/2,1/2}$* thresholds. At the top, an XAS spectrum is shown (dots) where the excitation energies for the RIXS spectra are indicated by the arrows aimed at the main peaks and satellite structures. A calculated isotropic XAS spectrum with the corresponding multiplets including all symmetries is also included (full curve). The final states of the NiO XAS spectrum are the same as the intermediate states in the RIXS process and are well understood in terms of atomic transitions in a crystal field [29]. The main peaks in the XAS spectrum around ∼ 853.2 eV and ∼ 870.3 eV which are separated by the 2*p* spin-orbit splitting are made up of the crystal-field splitted $2p^53d^9$ configuration, hybridized with the $2p^53d^{10}\underline{L}^{-1}$ manifold. The RIXS spectra in the lower part of Fig. 1 are plotted on an emission photon energy scale, with excitation energies denoted by the letters A-H from 852.0 eV up to 916.5 eV. As observed, the spectral shape strongly depends on the photon energy.

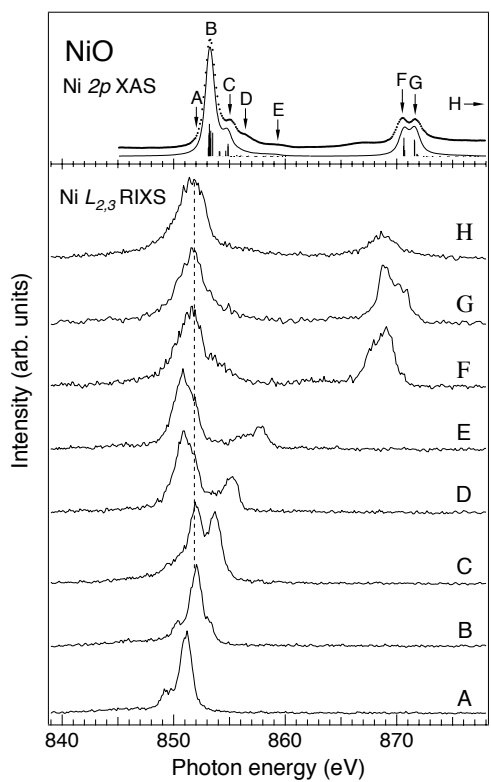

**Figure 1:** At the top is an XAS spectrum (dots) of NiO measured at the Ni *2p* edges compared to a calculated XAS spectrum (full curve) broadened with a Lorenzian (0.6 eV and 0.8 eV FWHM for the *2p$_{3/2}$* and *2p$_{1/2}$* thresholds, respectively) and a Gaussian profile of 0.4 eV. Below are measured $L_{2,3}$ RIXS spectra (full lines) on a photon energy scale excited at 852.0, 853.3, 855.1, 856.5, 859.3, 870.5, 871.7 and, 916.5 eV, denoted by the letters A-H, respectively.

The fluorescence data basically contain peak structures of three different categories: 1) recombination due to elastic $2p^53d^9 \rightarrow 3d^8$ transitions back to the ground state, also known as Rayleigh scattering, 2) the resonating loss structures due to *d−d* excitations and charge-transfer excitations of Raman scattering, and 3) normal X-ray emission lines. In order to minimize the elastic contribution at threshold excitation, the emitted photons were recorded at an angle, perpendicular to the direction of the incident photons, with the polarization vector parallel to the horizontal scattering plane. The elastic and Raman scattering contributions disperse on the photon energy scale while other structures due to normal emission do not reveal any major energy shifts at all. In spectrum F, excited at the $2p_{1/2}$ threshold, the main $L_3$ emission peak at ~ 851.8 eV is due to normal X-ray fluorescence at constant photon energy as indicated by the dashed vertical line. Since the ground state is dominated by the $3d^8$ configuration, the normal $L_{2,3}$ fluorescence intensity is mainly due to decays of the $2p^5 3d^8 \rightarrow 3d^7$ transitions, where the final states have one electron less than in the ground state. When the excitation energy is tuned to the charge-transfer satellite region in the absorption spectrum at D and E, it gives rise to intense emission lines at ~ 850.8 eV. The energy position of these lines is ~ 1.0 eV lower than for the normal emission line. This distinct energy shift shows that the origin of this emission line is not normal fluorescence but rather due to charge-transfer excitations as will be discussed below.

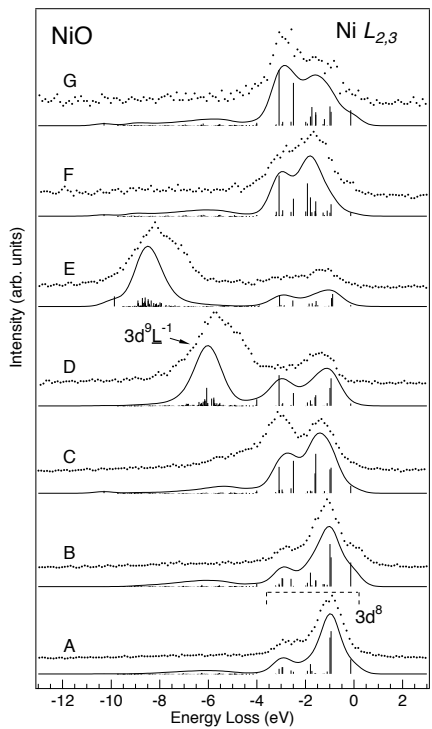

**Figure 2:** Calculated RIXS spectra (full lines) of NiO compared to the experimental spectra in Fig. 1 (dots) normalized to the same peak heights. The letters corresponds to the same excitation energies as in Fig. 1.

Figure 2 shows the $L_{2,3}$ RIXS data (dots) together with the results of SIAM calculations plotted as a function of the energy-loss or Raman shift. The energy loss is derived from the RIXS spectra by subtracting the incident photon energy from the energy of the emitted photons. In order to enhance the spectral shape modifications, the spectra are normalized to the same peak heights. The calculations were made in the same geometry as the experimental one (see section II). The letters A-H denote the same excitation energies as in Fig. 1. As observed, the Raman scattering calculations are generally in good agreement with the experimental data although contributions from normal fluorescence are not included in the theory. The peak structure at 0 eV energy loss corresponds to the elastic recombination peak back to the $^3A_{2g}$-derived ground state. Other prominent features corresponding to Raman scattering appears at ~ 1.1 eV, ~ 1.6 eV and ~ 2.8 eV. In addition, in the energy region 4-9 eV, there is a broad satellite structure with relatively low intensity. The Raman scattering peaks at ~ 1.1 eV, ~ 1.6 eV and ~ 2.8 eV, which stay at constant energy loss, essentially identified in all spectra with different intensities, are due to the ligand field splitting of the $3d^8$ final states. In Table II, the assignments of the loss structures are compared to those of optical absorption and EELS data. The lowest and most intense loss structure at ~ 1.1 eV above the ground state, most clearly observed in spectrum B, is due to the

$^3A_{2g} \rightarrow ^3T_{2g}$ triplet-triplet transition. The assignment of the ~ 1.6 eV loss structure observed as a shoulder in spectra B and C, is more difficult. We attribute this feature to a superposition of the triplet-triplet $^3A_{2g} \rightarrow ^3T_{1g}$ and the triplet-singlet $^3A_{2g} \rightarrow ^1E_g$ spin-flip transition in agreement with EELS measurements [9, 10].

**Table 1:** The parameter values used in the Anderson impurity model calculations. κ is the scaling factor for the Slater integrals, Δ is the energy difference between the gravity centers of the $3d^7$ and the $3d^8\underline{L}^{-1}$ configurations, $V_{eg1}$ and $V_{eg2}$ are the hybridization strengths for the $e_g$ orbitals in the ground and core-excited states, respectively. $W$ is the O $2p$ bandwidth, $Q$ is the core-hole potential, $U$ is the on-site Coulomb interaction between the localized $3d$ electrons and $10Dq$ is the crystal-field splitting. The super-exchange field was applied along the z-axis. All values, except for κ are in units of eV:s.

| κ | Δ | $V_{eg_1}$ | $V_{eg_2}$ | $W$ | $Q-U$ | $10Dq$ | Ex.field |
|---|---|---|---|---|---|---|---|
| 0.8 | 3.5 | 2.2 | 1.8 | 5.0 | 1.0 | 0.5 | 0.15 |

In spectrum B, the prominent peak at ~ 2.8 eV is assigned mainly to the $^3A_{2g} \rightarrow ^1T_{2g}$ triplet-singlet spin-flip transition. However, the detailed assignment of the measured energy-loss peak in the energy region 2-4 eV is complicated and there is also some influence of the $^3A_{2g} \rightarrow ^3T_{1g}$ triplet-triplet transition. The calculations of the spectral RIXS profiles reveal the sensitivity to the superexchange field which gives rise to a spectral weight transfer towards lower loss energies as it is increased [30]. Thus, the magnitude of the superexchange energy of 0.15 eV needed to reproduce the experimental spectra gives important information about the degree of covalent bonding which gives rise to the antiferromagnetic alignment of the ions in the crystal field. The intense dispersing lines with loss energies of ~ 5.7 eV and ~ 8.2 eV in spectra D and E are reproduced by the model calculations as resonances of ligand $2p \rightarrow$ metal $3d$ charge-transfer excitations to $3d^9\underline{L}^{-1}$ final states. A comparison between Figures 1 and 2 shows that these final states remain at constant photon energy on the emission energy scale and disperse on the energy-loss scale. Similar resonantly enhanced energy-loss structures as those observed in spectra D-E have been assigned to have charge-transfer origin in RIXS spectra of rare earths compounds [31].

It is important to note that the relative energy positions of the peak structures in RIXS, OAS and EELS are different from the binding energies in RPES. For valence-band RPES, excited at the $2p$ thresholds [11], the main $3d$ peak structure is observed at the lowest binding energy with an intensity maximum at ~ 3.5 eV. In the localized picture, this peak has been interpreted as a $3d^8\underline{L}^{-1}$ final state. The $3d$ peak is followed by the oxygen $2p$ band at ~ 3-8 eV binding energy. At ~ 6-12 eV binding energy, an even broader peak structure is due to the so-called satellite. In contrast to the main $3d^8\underline{L}^{-1}$ line, the satellite show a strong resonantly enhanced behaviour and has been explained as a superposition of Autoionization and direct photoemission processes leading to the same $3d^7$ final state [2].

**Table 2:** Ground state and final state energies of *d−d* transitions in NiO(100) measured by RIXS, optical absorption [7, 8] and EELS [9, 10]. All values are in eV:s.

| Symmetry | RIXS | Opt. abs. | EELS |
|---|---|---|---|
| $^3A_{2g}$ | 0 | 0 | 0 |
| $^3T_{2g}$ | 1.1 | 1.08-1.13 | 1.12 |
| $^1E_g$ | 1.6 | 1.73-1.75 | 1.60 |
| $^3T_{1g}$ | 1.6 | 1.88-1.95 | 1.70 |
| $^1T_{2g}$ | 2.8 | 2.73-2.75 | 2.75 |
| $^3T_{1g}$ | | 2.95 | 2.9 |
| $^1T_{1g}$ | | 3.26-3.52 | 3.1 |

The band gap of NiO has been estimated with the combination of RPES and inverse photoemission spectroscopy (IPES) on the same sample with the same reference point. In IPES, probing the empty valence states, one electron is added to the ground state configuration. The main peak structure with the lowest binding energy is located at ∼ 3.9 eV interpreted in the localized picture as a $3d^9$ final state. The difference between the energy positions and intensities of the spectral structures in RIXS, RPES and IPES is generally a result of probing the systems with different numbers of electrons in the final states in comparison to the ground state. The $3d^8\underline{L}^{-1}$ final state in RPES has a sizable binding energy indicating that this structure is not close to the ground state but is an excited state manifested by the strong correlation effect. The RPES $3d^8\underline{L}^{-1}$ final state thus appear at relatively larger binding energies than the $3d^8$ energy-loss structures observed in RIXS and related spectroscopies (see Table II). Since all RPES final states in general have one electron removed, they represent excited states not directly reflecting the electronic structure of the ground state. The size of the experimental band gap (3.9-4.3 eV) between the half heights of the RPES $3d^8\underline{L}^{-1}$ and IPES $3d^9$ final states therefore contains a certain amount of personal judgement [2, 3]. The RIXS excitation-deexcitation process, on the other hand, is not only element selective but is also charge neutral leading to localized $3d^8$ final states due to *d−d*-excitations as well as screened $3d^9\underline{L}^{-1}$ charge-transfer final states directly probing the ground as well as the low-energy excited states. The RIXS technique is here shown to be very sensitive for detecting the localized *d−d* excitations from the ground state as well as the important charge-transfer excitations in strongly correlated materials like NiO.

# 5 Summary

The electronic structure of NiO has been measured at the Ni 2*p* absorption thresholds by resonant soft X-ray Raman scattering. The spectral energy-loss features due to *d−d* and charge-transfer excitations and normal fluorescence are identified by changing the incoming photon energy. Spectral simulations within the single impurity Anderson model are consistent with the experimental data implying high sensitivity to the crystal field and superexchange interactions. At threshold excitation, crystal-field splitted triplet-triplet and spin-flip triplet-singlet *d−d* excitations of the $3d^8$ configuration are identified and compared to other spectroscopical techniques. Pronounced energy-loss structures due to charge-transfer excitations to the $3d^9\underline{L}^{-1}$ final state are also found.

# 6 Acknowledgments

This work was supported by the Swedish Natural Science Research Council (NFR), the Göran Gustafsson Foundation for Research in Natural Sciences and Medicine and the Swedish Foundation for International Cooperation in Research and Higher Education (STINT). The authors would like to thank Prof. A. Kotani for useful discussions.